\documentclass[aps,prb,floatfix,twocolumn,showpacs,reprint,superscriptaddress]{revtex4-1}
\usepackage{graphicx}
\usepackage{mathrsfs}
\usepackage{bm}
\usepackage{amsmath}
\usepackage{dcolumn}
\usepackage{epstopdf}
\usepackage{dsfont}
\usepackage{amssymb}
\usepackage{tabularx}
\usepackage{array}
\usepackage{color}
\usepackage[colorlinks=true, letterpaper=true, pdfstartview=FitV, linkcolor=blue, citecolor=blue, urlcolor=blue]{hyperref}

\begin{document}
\title{Valley-polarized quantum anomalous Hall phase and disorder induced valley-filtered chiral edge channels}

\author{Hui Pan}
\affiliation{Department of Physics, Beihang University, Beijing 100191, China}

\author{Xin Li}
\affiliation{Department of Physics, Beihang University, Beijing 100191, China}

\author{Hua Jiang}
\affiliation{College of Physics, Optoelectronics and Energy, Soochow University, Suzhou 215006, China}

\author{Yugui Yao}
\affiliation{School of Physics, Beijing Institute of Technology, Beijing 100081, China}

\author{Shengyuan A. Yang}
\affiliation{Engineering Product Development, Singapore University of Technology and Design, Singapore 138682, Singapore}

\begin{abstract}
We investigate the topological and transport properties of the recently discovered valley-polarized quantum anomalous Hall (VQAH) phase. In single layer, the phase is realized through the competition between two types of spin-orbit coupling, which breaks the symmetry between the two valleys. We show that the topological phase transition from conventional quantum anomalous Hall phase to the VQAH phase is due to the change of topological charges with the generation of additional skyrmions in the real spin texture, when the band gap closes and reopens at one of the valleys.
In the presence of short range disorders, pairs of the gapless edge channels (one from each valley in a pair) would be destroyed due to intervalley scattering.
However, we discover that in an extended range of moderate scattering strength, the transport through the system is quantized and fully valley-polarized, i.e. the system is equivalent to a quantum anomalous Hall system with valley-filtered chiral edge channels.
We further show that with additional layer degree of freedom, much richer phase diagram could be realized with multiple VQAH phases. For a bilayer system, we demonstrate that topological phase transitions could be controlled by the interlayer bias potential.
\end{abstract}


\maketitle

\section{Introduction}

Gapless one-dimensional (1D) edge channels are intriguing physical objects which are usually associated with nontrivial topological phases of two-dimensional (2D) systems. For example, the precise quantization of Hall plateaus in quantum Hall effect is tied with the dissipationless chiral edge channels and is related to a bulk topological invariant known as the Chern number (or TKNN invariant).\cite{Laughlin,Thouless} It was later realized that the existence of such edge channels in fact does not necessarily require the orbital effects of external magnetic field. Instead, it could arise from the combined effects of spin polarization (e.g. due to magnetic ordering) and spin-orbit coupling (SOC).\cite{hald,onod,qi,qi2008,wang2014} This topological phase, known as the quantum anomalous Hall (QAH) phase, has been sought for more than 20 years. Hence its first realization in magnetically doped topological insulator thin films has attracted significant attention and research activities recently.\cite{YuR,ChangCZ,chec,kou,JiangH}

It is possible that these edge channels may carry additional flavors. For example, in quantum spin Hall effect, the chirality of the edge channel is tied with its spin, hence on each edge, there is a Kramers pair of counter-propagating spin-polarized edge channels, which are protected by time reversal symmetry.\cite{KaneCL1,ZhangSC1} When the band structure has multiple energy extremas, carriers could have another type of flavor, valley. Similar to spintronics, it was proposed that this valley degree of freedom may also be utilized for information processing, leading to the concept of valleytronics.\cite{ryce2007,guna2006,xiao2007,yao2008,zhu2012,xu2014,KFMak} It has been shown that there could be topological charges associated with the valleys and it is possible to realize 1D channels that carry specific valley indices.\cite{Morpurgo,YaoW2,QiaoZH1,QiaoZH2,MacDonald2,YKim} However, in these previous studies, the numbers of 1D channels in each valley are balanced, due to the presence of either time reversal symmetry or inversion symmetry.

In a recent work, we demonstrated that by breaking both time reversal symmetry and inversion symmetry, it is possible to achieve a novel topological phase, the valley-polarized quantum anomalous Hall (VQAH) phase.\cite{PanH} The hallmark of this phase is that at system edges where valleys can be distinguished, there exist unbalanced numbers of counter-propagating chiral edge channels associated with the two valleys. This imbalance automatically indicates that the system is in a QAH phase. The additional valley features of the edge channels are manifestations of the unbalanced valley topological charges in the bulk. Therefore such phase is characterized by two bulk topological invariants: the total Chern number $\mathcal{C}$ and the valley Chern number $\mathcal{C}_v$. We have found that such a novel phase could be realized due to the competition between two types of SOCs in a low buckled honeycomb lattice model which may describe 2D materials such as silicene or germanene.\cite{Lalmi,LiuCC1,LiuCC2,Vogt,Fleurence,Ezawa1,ChenL,Tsai}

Due to length restrictions, several important physical aspects of the VQAH phase were not exposed in the previous work. In the present paper, we would address these details. More importantly, we greatly extend our previous work by investigating the disorder effects on VQAH phase and the valley-polarized topological phases in a bilayer system. The valley polarized channels are robust against smooth-varying scattering potentials, expected from the large momentum separation between the two valleys. Through explicit transport calculations, we show that for short range scatterers, although the intervalley scattering would destroy pairs of counter-propagating valley channels, remarkably, the remaining $|\mathcal{C}|$ channels can still retain their valley character in transport. As a result, each edge could serve as a perfect valley filter with chiral edge channels for one specific valley. We show that this happens for an extended window of intermediate scattering strength. Furthermore, for a bilayer system formed by stacking two single layers, we find that the resulting properties are not simply superposition of the two. In fact, the bilayer system exhibits a much richer phase diagram including several VQAH phases with different $(\mathcal{C},\mathcal{C}_v)$ invariants. The topological phase transitions between these phases could be more easily controlled by tuning the interlayer bias potential.

Our paper is organized as following. In Sec. II, we discuss the VQAH phase in a single layer lattice model. By decomposing the topological charges into real spin and pseudospin sectors, we show that the topological phase transition to VQAH phase is accompanied with the change of real spin topological charge in one valley. In Sec. III, we study the effects of short range scattering on the VQAH phase based on a two-terminal transport calculation and show that in a window of intermediate scattering strength, only one chiral edge channel with $K'$ valley character is left. In Sec. IV, we investigate the rich topological phases in a bilayer system and demonstrate the tunability of the phases through interlayer bias and exchange field strength. Finally, we give our conclusion and summarize our results in Sec. V.

\section{Valley-polarized QAH phase in single layer system}

The VQAH phase was first discovered in a lattice model defined on a low-buckled single layer honeycomb lattice.\cite{PanH} The tight-binding Hamiltonian is written as\cite{LiuCC2}
\begin{equation}\begin{split}\label{H1}
H=&-t\sum_{\langle ij\rangle \alpha}c^\dagger_{i\alpha}c_{j\alpha}+ i t_\text{SO}\sum_{\langle\langle ij \rangle\rangle \alpha\beta}
 \nu_{ij}c^\dagger_{i\alpha}{s}^{z}_{\alpha\beta}c_{j\beta}\\
 &-i t_{\text{R}_2}\sum_{\langle\langle ij \rangle\rangle \alpha\beta}\mu_{ij}
 c^\dagger_{i\alpha}(\bm{s} \times \hat{\bm{d}}_{ij})^{z}_{\alpha\beta} c_{j\beta}\\
 &+it_{\text{R}_1}\sum_{\langle ij \rangle \alpha\beta}c^\dagger_{i\alpha}
 (\bm{s} \times \hat{\bm{d}}_{ij})^{z}_{\alpha\beta} c_{j\beta}+M\sum_{i\alpha\beta}c^\dagger_{i\alpha}{s}^{z}_{\alpha\beta}c_{i\beta}.
\end{split}
\end{equation}
Here $c^\dagger_{i\alpha}$($c_{i\alpha}$) is a creation (annihilation) operator for an electron with spin $\alpha$ at site $i$.  The summation with $\langle...\rangle$ ($\langle\langle...\rangle\rangle$) runs over all nearest (next-nearest) neighbor sites. The $s$'s are the Pauli matrices for real spin degree of freedom. For the right hand side, the first term is the usual nearest neighbor hopping term. The second term is the so-called intrinsic SOC term involving the next-nearest neighbor hopping,\cite{kane2005a,kane2005b} $\nu_{ij}=+1 (-1)$ if the electron makes a left (right) turn in going from site $j$ to site $i$ along the nearest-neighbor bonds. The third and fourth terms are the intrinsic and extrinsic Rashba SOC terms respectively. $\hat{\bm{d}}_{ij}$ is the unit vector pointing from site $j$ to $i$, and $\mu_{ij}=\pm 1$ depending on the $AB$-sublattice.
The last term represents an exchange coupling. The various $t$'s and $M$ denote the strengths of the terms.

This model was first derived in the study of low energy physics of silicene.\cite{LiuCC2} The various SOC terms are the symmetry-allowed terms for the low buckled honeycomb lattice structure.\cite{YangSY1} The exchange term breaks the time reversal symmetry, which is necessary for the realization of QAH phase.\cite{hald,PanH2} As discussed in the previous work,\cite{PanH} the VQAH phase results from the competition between the intrinsic and extrinsic Rashba SOC terms, i.e. the third term and the fourth term in Eq.(\ref{H1}). The intrinsic Rashba term is due to the mirror symmetry breaking of the 2D plane from the lattice buckling.\cite{LiuCC2} The extrinsic Rashba term further breaks the inversion symmetry and it can result from a perpendicular electric field or from a substrate. To simplify the analysis, in the following we shall focus on these two SOC terms and neglect the intrinsic SOC term. The effects of the intrinsic SOC term and possible sublattice symmetry breaking term will be discussed later in the paper.

\begin{figure}
  \includegraphics[width=8.8cm]{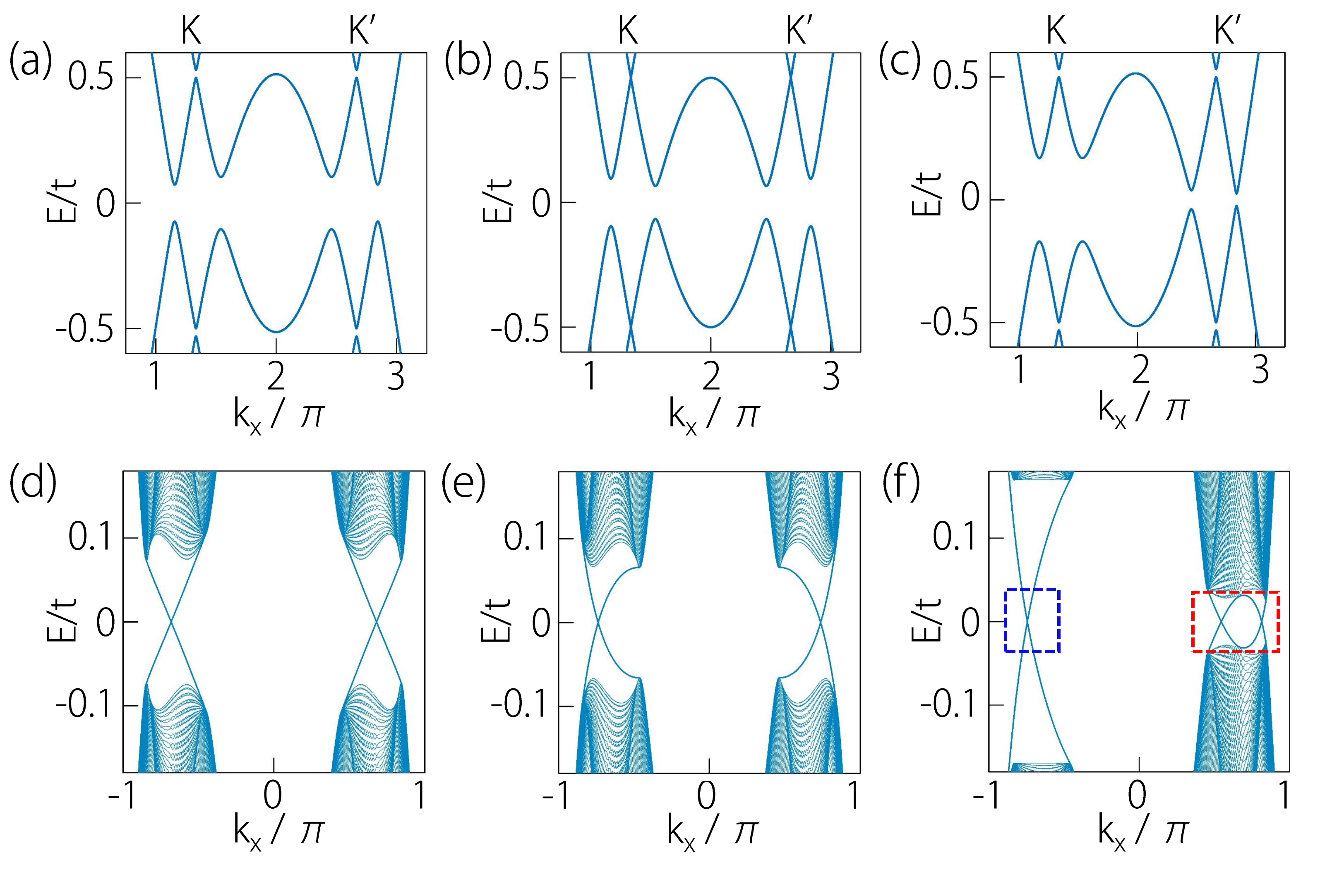}
  \caption{\label{FIG:EkQAH} (color online) Upper panel: bulk band structure along the line of $k_y=0$. Lower panel: the corresponding energy spectra for a zigzag edged ribbon with a width of 400 atomic sites. (a, d) with only extrinsic Rashba SOC $t_{\text{R}_1}=0.06$ and no intrinsic Rashba SOC; (b, e) with only intrinsic Rashba SOC $t_{\text{R}_2}=0.1$ and no extrinsic Rashba SOC; (c, f) when both extrinsic and intrinsic Rashba SOCs are present, $t_{\text{R}_1}=0.06$ and $t_{\text{R}_2}=0.1$. Other model parameters are set as $t=1$, and $M=0.5$.}
\end{figure}

First, we examine the properties of the model when either intrinsic Rashba or extrinsic Rashba term is present, but not both. Figure~\ref{FIG:EkQAH}(a) shows the bulk energy spectrum near the band gap along $k_x$ direction when only extrinsic Rashba SOC is present, and Fig.~\ref{FIG:EkQAH}(d) shows the corresponding energy spectrum for a ribbon with zig-zag edge termination. The results for the case with only intrinsic Rashba SOC are shown in Fig.~\ref{FIG:EkQAH}(b) and \ref{FIG:EkQAH}(e). One observes that for both cases, the system is in insulating state with a finite band gap. In the results for ribbons, four gapless chiral edge states can be identified in the band gap. From their wave functions, it is easily checked that on each edge, there are two edge states propagating in the same direction, indicating a QAH phase with $\mathcal{C}=2$. The Chern number can be directly calculated from the bulk band structure using the formula\cite{Thouless}
\begin{equation}\label{EQ:CN}
\mathcal{C}=\frac{1}{2\pi}\sum_{n\in\text{occ.}}\int_\text{BZ}d^2\bm{k}\,\mathit\Omega_n,
\end{equation}
where the integration is over the Brillouin zone and the summation is over all occupied valence bands. $\mathit\Omega_n$ is the momentum-space Berry curvature for the $n$-th band
\begin{equation}
\mathit\Omega_n(\bm{k})=-\sum_{n'\neq n}
\frac{ 2\mathrm{Im}\langle\psi_{n\bm k}|v_x|\psi_{n'\bm k}\rangle
 \langle\psi_{n'\bm k}|v_y|\psi_{n\bm k}\rangle }{ (\varepsilon_{n'\bm k}-\varepsilon_{n\bm k})^{2} },
\end{equation}
where $v_{x(y)}$ is the velocity operator and $|\psi_{n\bm k}\rangle$ is the Bloch eigenstate with eigen-energy $\varepsilon_{n\bm k}$.
The magnitude of Berry curvature is usually peaked at avoided band crossings where the gap is small.\cite{naga2010} For a system with multiple valleys, such as the case here with two valleys $K$ and $K'$, Berry curvature will be concentrated around the valley centers.\cite{PanH} This allows us to define a topological charge associated with each valley by integrating the Berry curvature over the neighborhood of each valley as in Eq.(\ref{EQ:CN}).\cite{volovik,Morpurgo,YaoW2} We denote the results by $\mathcal{C}_K$ and $\mathcal{C}_{K'}$. They represent the contribution to the total Chern number from each valley, and their difference $\mathcal{C}_v=\mathcal{C}_K-\mathcal{C}_{K'}$ is called the valley Chern number.
Note that the concept of valley as well as the valley topological numbers are well-defined only when the low energy regions are well separated in the reciprocal space. This condition is ensured in our following calculations. In the studied parameter range, the various SOCs are small perturbations compared with the nearest-neighbor hopping which is the largest energy scale.

Straightforward calculations using the present model confirm that for both cases, the system is in the same QAH phase with $\mathcal{C}=2$, and $\mathcal{C}_v=0$ showing that contribution from the two valleys are equal ($\mathcal{C}_K=\mathcal{C}_{K'}=1$). Indeed, on one edge, each valley contributes one chiral edge channel propagating in the same direction.

\begin{figure}
  \includegraphics[width=8.6cm]{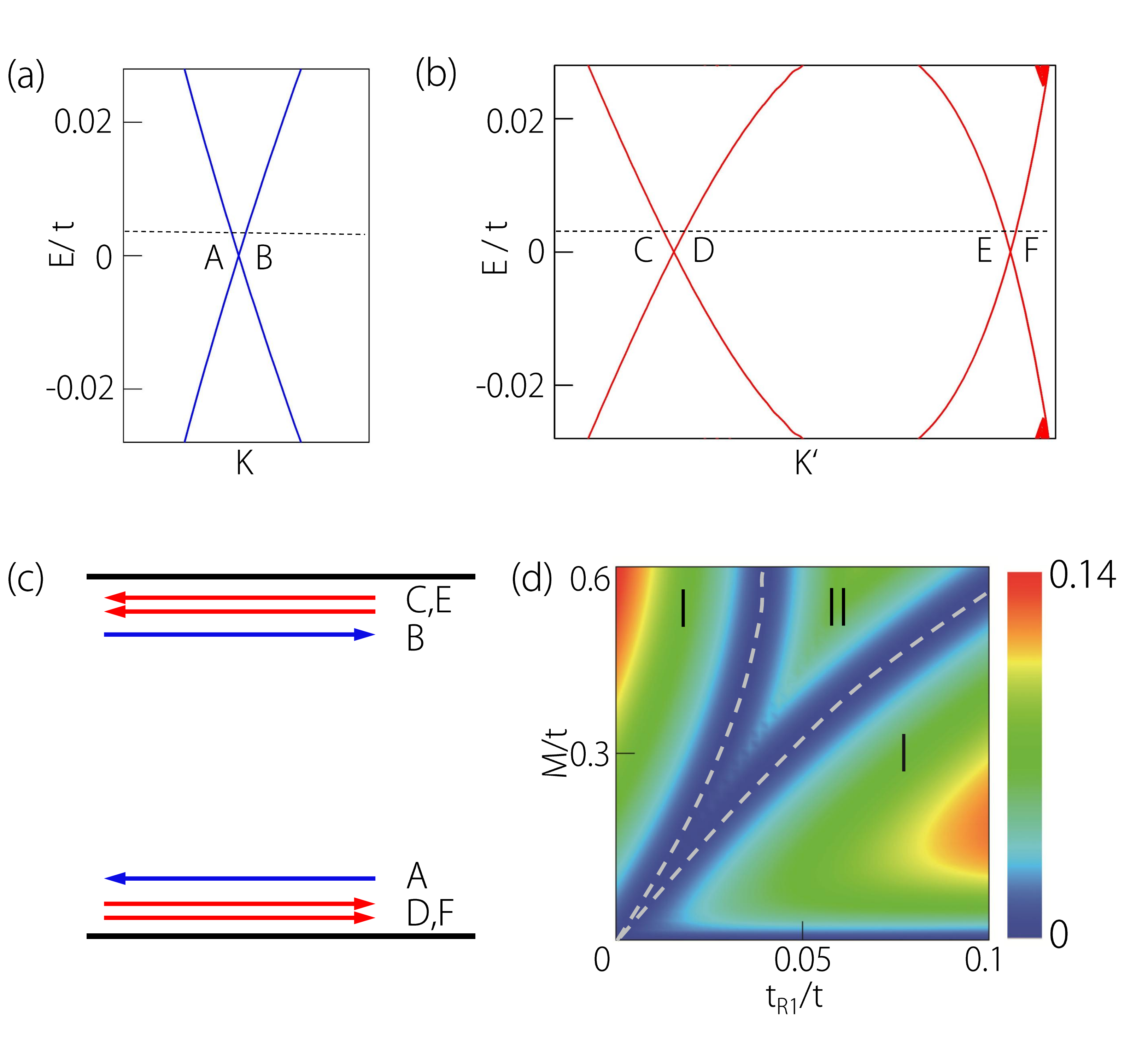}
  \caption{\label{FIG:EdgeVQAH} (color online) (a) and (b) are the enlarged spectra showing the gapless edge states (a) in $K$ valley and (b) in $K'$ valley, corresponding to the boxes in Fig~\ref{FIG:EkQAH}(f). (c) Schematic figure of the spatial distribution of the edge channels in the ribbon. The colors label the edge states at different valleys. (d) Phase diagram as a function of $t_{\text{R}_1}$ and $M$. The dashed lines are the phase boundaries where the bulk gap closes. Phase I is the conventional QAH phase, while Phase II is the VQAH phase.}
\end{figure}

The situation becomes quite different when both SOCs are present. Starting from a fixed intrinsic Rashba SOC as in Fig.~\ref{FIG:EkQAH}(b, e), gradually increasing the strength of extrinsic Rashba SOC, it has been shown that the gap at $K$ valley remains open but the gap at $K'$ valley closes and reopens, leading to a topological phase transition to the VQAH phase with $\mathcal{C}_v\neq 0$. As shown in Fig.~\ref{FIG:EkQAH}(c) and \ref{FIG:EkQAH}(f), in VQAH phase, the two valleys become asymmetric. In $K$ valley, there are still two gapless edge states, but in $K'$ valley there are four. These states can be more clearly seen in the zoom-in images in Fig.~\ref{FIG:EdgeVQAH}(a) and \ref{FIG:EdgeVQAH}(b). In Fig.~\ref{FIG:EdgeVQAH}(c), we schematically plot the spatial distribution of these edge states in the ribbon geometry. One notes that on each edge, there are two chiral channels from $K'$ valley propagating in opposite direction to only one channel from $K$ valley. Calculation of the topological charge shows that $\mathcal{C}_K$ in this case is still 1, but $\mathcal{C}_{K'}$ changes from 1 to $-2$, which is consistent with the doubling of the $K'$ valley edge channels and the reversed chirality. In such a VQAH phase, on the edge, there is an imbalance between channels from different valleys, and in the bulk, it is characterized by both a nonzero $\mathcal{C}$ and a nonzero $\mathcal{C}_v$. In the present case, we have $(\mathcal{C}=-1,\mathcal{C}_v=3)$.

By looking at the figures in Fig.~\ref{FIG:EkQAH}, one notices that for the cases with either one Rashba SOC, the energy spectra at $K$ and $K'$ valleys are symmetric. But when both SOCs are present, the symmetry of the spectra is broken. To understand this, we expand the model around the $K$ and $K'$ points to obtain the low energy effective Hamiltonian. The corresponding forms of the kinetic energy term, the extrinsic Rashba term, the intrinsic Rashba term, and the exchange coupling term are
\begin{subequations}
\begin{align}
&\mathcal{H}_{0}(\bm{k}) = v(\tau_{z}\sigma_x k_x + \sigma_y k_y), \\
&\mathcal{H}_{\text{R}_1}(\bm{k}) = \lambda_{\text{R}_1}(\tau_{z}\sigma_x s_y - \sigma_y s_x),  \\
&\mathcal{H}_{\text{R}_2}(\bm{k}) = \lambda_{\text{R}_2}\sigma_z (k_y s_x - k_x s_y), \\
&\mathcal{H}_{M}(\bm{k}) = M s_z,
\end{align}
\end{subequations}
where $\tau_z=\pm 1$ refers to $K$ and $K'$ valleys, $\sigma$'s are Pauli matrices representing the $AB$-sublattice pseudospin degree of freedom, the coupling strengths in these terms are related to the parameters in Eq.(\ref{H1}) by $v=\sqrt{3}t/2$, $\lambda_{\text{R}_1}=3t_{\text{R}_1}/2$, and
$\lambda_{\text{R}_2}=3t_{\text{R}_2}/2$. When the extrinsic Rashba SOC is absent, i.e. $\lambda_{\text{R}_1}=0$, as in Fig.~\ref{FIG:EkQAH}(b, e), the remaining terms all have inversion symmetry, $\mathcal{P}=\sigma_x$, such that $\mathcal{P}\mathcal{H}(\bm k)\mathcal{P}^{-1}=\mathcal{H}(-\bm k)$, meaning that the spectra at the two valleys must be symmetric under inversion. When extrinsic Rashba term is present, the inversion symmetry is broken. However, in the absence of intrinsic Rashba term, the low energy model has another symmetry $\mathcal{Q}=\sigma_x s_z$ which is an inversion with an additional spin rotation, such that $\mathcal{Q}\mathcal{H}(\bm k)\mathcal{Q}^{-1}=\mathcal{H}(-\bm k)$. Therefore the spectra in Fig.~\ref{FIG:EkQAH}(a, d) also exhibit similar symmetric feature. We emphasize that $\mathcal{Q}$ is not an intrinsic symmetry for the crystal, it is an emergent symmetry only for the low energy model. Finally when both Rashba terms are present, the two symmetries $\mathcal{P}$ and $\mathcal{Q}$ are both broken. The spectra at the two valleys become asymmetric, as observed in Fig.~\ref{FIG:EkQAH}(c, f). The asymmetry between the two valleys is a necessary condition for realizing the VQAH state.

The topological phase transitions from conventional QAH phase to VQAH phase can be realized by tuning the model parameters. In Fig.~\ref{FIG:EdgeVQAH}(d), we show the phase diagram in the $(t_{\text{R}_1}$-$M)$ plane at a fixed intrinsic Rashba strength. It can be seen that the VQAH phase has an extended parameter range in the phase diagram (region II). On each side of VQAH phase, it is the usual QAH phase with $(\mathcal{C}=2, \mathcal{C}_v=0)$. The color map shows the size of the band gap. One observes that the topological phase transitions are accompanied with the gap closing and reopening processes, as usually mentioned in the study of topological insulators.\cite{ZhangSC1} However, in the present model, such gap closing happens only at one valley (the $K'$ valley), hence the topological charge is only changed at that valley, leading to the valley-polarized feature.

To gain a better understanding of the change in topological charge between conventional QAH and VQAH phases, we decompose $\mathcal{C}_K$ and $\mathcal{C}_{K'}$ at each valley in terms of contributions from real spin ($\bm s$) as well as sublattice pseudospin ($\bm \sigma$) degree of freedom. We calculate the winding number of the spin (pseudospin) textures using the formula\cite{qi2008}
\begin{equation}
n=\frac{1}{4\pi}\int\int dk_xdk_y (\partial_{k_x}\hat{\bm h}
\times \partial_{k_y}\hat{\bm h})\cdot \hat{\bm h},
\end{equation}
where the unit vector $\hat{\bm h}(\bm k)$ is the spin (pseudospin) polarization vector at $\bm k$.

For the conventional QAH phase with either extrinsic or intrinsic Rashba SOC, as in Fig.~\ref{FIG:EkQAH}(a, b), we find that the band-resolved topological charges carried
by the real spin or pseudospin for valleys $K$ and $K'$ are
\begin{equation}\label{QAHn}
\begin{split}
n_{K,1s}= n_{K',1s}\approx 0; \\
n_{K,2s}= n_{K',2s}\approx 1; \\
n_{K,1\sigma}= n_{K',1\sigma}\approx 0.5; \\
n_{K,2\sigma}= n_{K',2\sigma}\approx -0.5.
\end{split}
\end{equation}
Here subscripts 1 and 2 refer to the two valence bands with band 2 close to the gap.
One observes that the topological charges are symmetric between the two valleys. The topological charges of pseudospin from the $1$st and $2$nd valence bands are respectively $0.5$ and $-0.5$ for each valley, hence cancelling each other. The net contribution is from the real spin in the $2$nd valence band which is $1$ for each valley, with the texture corresponding to one skyrmion. Therefore we have $\mathcal{C}_{K}=\mathcal{C}_{K'}=1$ and  $(\mathcal{C}=2,\mathcal{C}_{v}=0)$, which are consistent with previous calculations.

In the VQAH phase, as in Fig.~\ref{FIG:EkQAH}(c), straightforward calculation shows that
\begin{equation}
\begin{split}
n_{K,1s}= n_{K',1s}\approx 0; \\
n_{K,2s}\approx 1; n_{K',2s}\approx -2; \\
n_{K,1\sigma}= n_{K',1\sigma}\approx 0.5; \\
n_{K,2\sigma}= n_{K',2\sigma}\approx -0.5.
\end{split}
\end{equation}
Comparing with the results for the QAH phase in Eq.(\ref{QAHn}), one can find that the topological charges associated with pseudospin texture remain the same and are still cancelled between the two valence bands. The difference is from the real spin in the $K'$ valley of the 2nd valence band. After the gap closing and reopening at the $K'$ valley, the topological charge $ n_{K',2s}$ changes from $1$ to $-2$.  This results in an imbalance of the total topological charges between the two valleys, leading to $\mathcal{C}_{K}=1$ and $\mathcal{C}_{K'}=-2$ with an imbalanced number of chiral edge channels associated with the two valleys.

\begin{figure}
  \includegraphics[width=8.4cm]{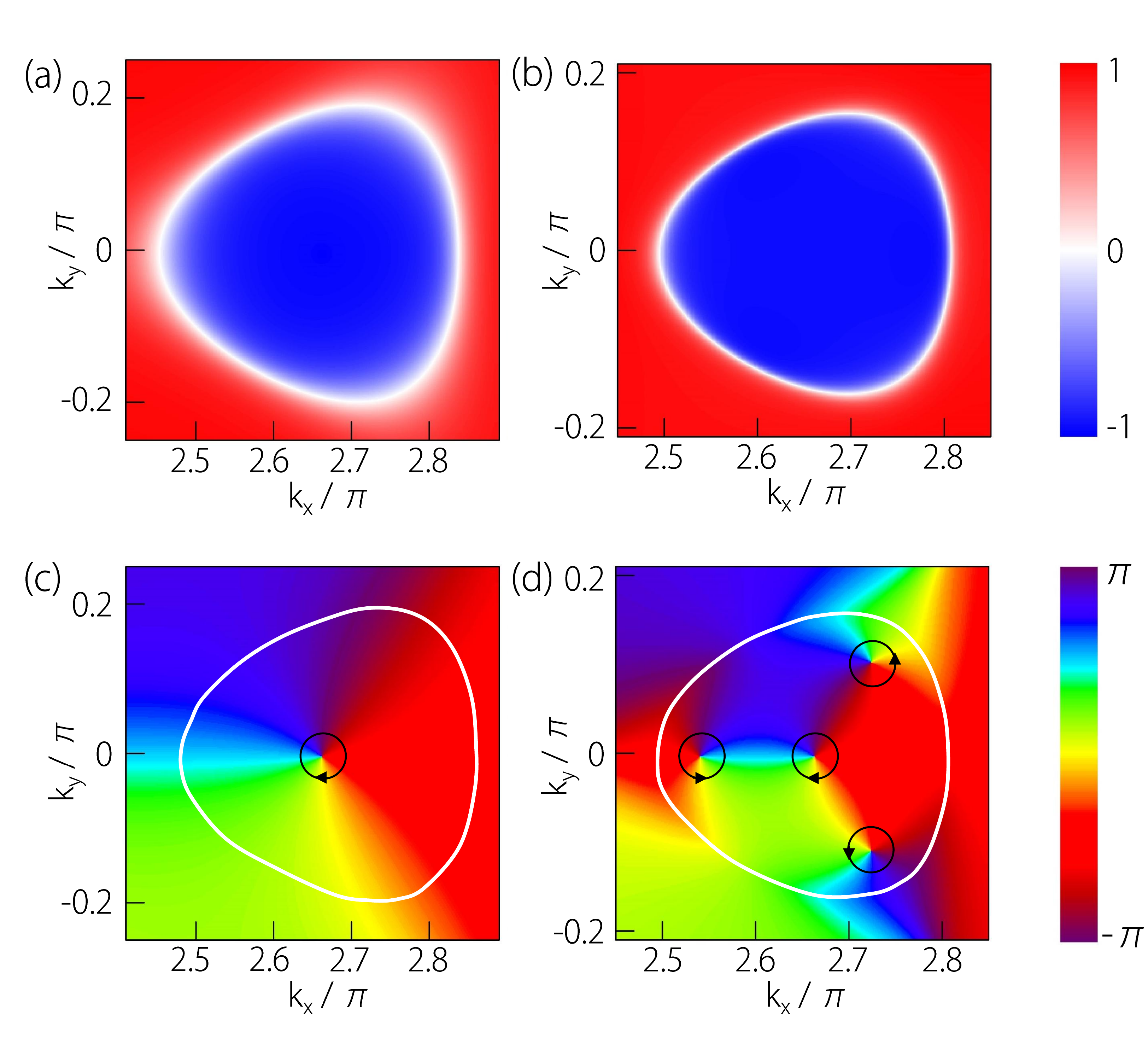}
  \caption{\label{FIG:SpinVector} (color online) Textures of real spin in the 2nd valence band of $K'$ valley. (a) Map of $\langle s_z\rangle$ component for the conventional QAH phase with ($t_{\text{R}_2}=0.1$, $t_{\text{R}_1}=0$). (b) Map of $\langle s_z\rangle$ for the VQAH phase with ($t_{\text{R}_2}=0.1$, $t_{\text{R}_1}=0.06$). (c) and (d) are the corresponding maps for the angle $\phi$ of the in-plane spin component. The white colored loop is the $\langle s_z\rangle=0$ boundary (as in (a) and (b)). The black circled arrows indicate the rotation direction of the angle $\phi$ around each vortex. }
\end{figure}

In order to visualize the change of $n_{K',2s}$ more clearly, in Fig.~\ref{FIG:SpinVector}, we plot the real spin textures in $k$-space for the 2nd valence band at $K'$ valley in the QAH phase and in the VQAH phase. Figure \ref{FIG:SpinVector}(a) and \ref{FIG:SpinVector}(b) show the $z$-component of real spin $\langle s_z\rangle$ before and after phase transition. One observes that near the valley center, $\langle s_z\rangle$ is negative and away from the center it changes to positive values. This feature remains the same across the phase boundary. Figure \ref{FIG:SpinVector}(c) and \ref{FIG:SpinVector}(d) show the azimuthal angle of the in-plane vector $(\langle s_x\rangle, \langle s_y\rangle)$ in the two phases. The winding number $n_{K',2s}$ can be visualized by counting the number of vortices of the phase winding. For the QAH phase, there is one vortex at the valley center, as seen in Fig.~\ref{FIG:SpinVector}(c), corresponding to $n_{K',2s}=1$. In the VQAH, in contrast, close to $K'$ point three new vortices appear around the places where the gap closes during the phase transition. Their winding directions are opposite to the one at the center, therefore leading to a total winding number $n_{K',2s}=-2$. It is these additional skyrmions generated at $K'$ valley in the gap closing and reopening process that are responsible for the realization of the VQAH phase.

\section{disorder induced valley-filtered chiral edge channels}

As we discussed in the previous section, in the VQAH phase of the present model, there exist valley-polarized chiral edge channels as shown in Fig.~\ref{FIG:EdgeVQAH}(c). Due to the large separation of the two valleys in $k$-space, the valley index is robust against smooth disorder potentials.\cite{ryce2007} In the presence of short-range disorder scattering, inter-valley scattering events would be important and would typically destroy the edge channels. Nevertheless, the VQAH phase is first of all a QAH phase characterized by a Chern number $\mathcal{C}=-1$. The topological protection of QAH phase is much stronger than that for the valley indices. Therefore, we can expect that at moderate short-range scattering strength, a pair of counter-propagating edge channels (one from $K$ and one from $K'$) should be destroyed, leaving only one chiral channel on each edge, as required by the $\mathcal{C}=-1$ constraint. An important question to ask is that whether the remaining channel still retains a valley character in terms of its transport property?

\begin{figure}
  \includegraphics[width=8.4cm]{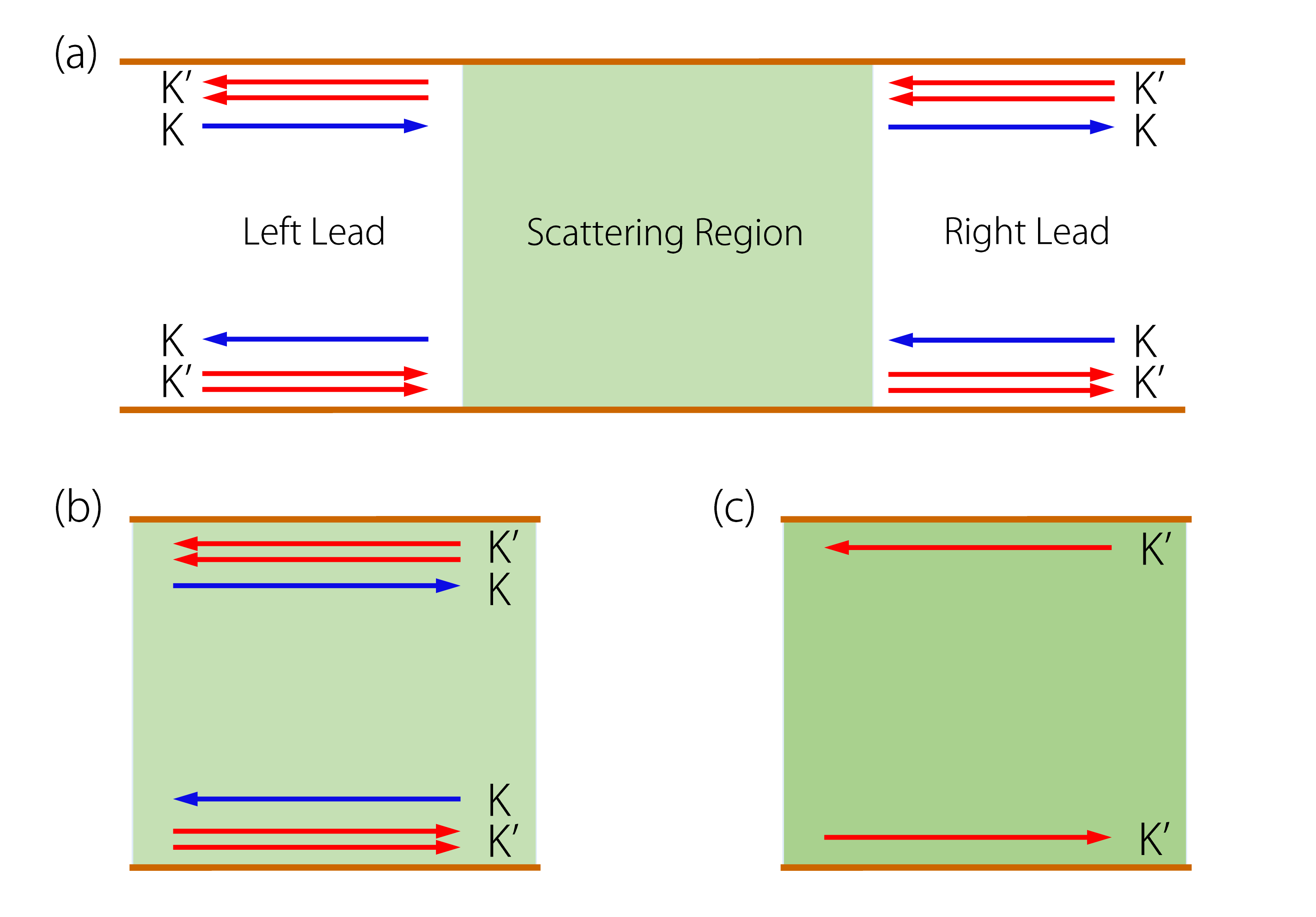}
  \caption{\label{FIG:EdgeState} (color online) Schematic figure showing the two-terminal setup for transport calculation. (a) A zigzag edged ribbon is divided into a left lead, a right lead, and a central scattering region. The propagating modes in the leads with their valley characters  are indicated. (b) Distribution of the edge modes in the scattering region without scattering or with only weak long range scattering. (c) With short range scattering, at moderate scattering strength, the transport property of the system is equivalent to a QAH system with one chiral edge channel in $K'$ valley.  The colors label the edge states at different valleys.}
\end{figure}

To address this question and to test the above physical picture, we carry out direct transport calculations on a standard two terminal structure.
As shown in Fig.~\ref{FIG:EdgeState}(a), the structure is divided into a left lead, a right lead, and a central scattering region. We take a zigzag-edged ribbon described by our model for the central region. The two leads are taken as semi-infinite. To eliminate the contact resistance and also for the analysis of the valley character of the transport channels, we model the leads with the same lattice model and with the same width as the central region. Short-range scatterers are confined in the central region and are modeled by random on-site disorder potentials with magnitude in the range $[-W/2, W/2]$. The parameter $W$ characterizes the disorder strength. We emphasize that due to disorder scattering, valley is no long a good quantum number in the central region. However, its valley transport property could be inferred from the transmission and reflection amplitudes of valley-polarized carriers from the leads. To this end, the two leads must have valley well-defined. Our setup resembles that of the original proposal of valley filter\cite{ryce2007} and we shall demonstrate that our system indeed acts as a perfect valley filter when moderate short-range scatterers are incorporated.

The two-terminal conductance can be calculated based on the Landauer-B\"{u}ttiker formula\cite{Datta}
\begin{equation}
G=\frac{e^2}{h}\texttt{Tr}[\Gamma_L G^r \Gamma_R G^a],
\end{equation}
where $G^{r,a}$ are the retarded and advanced Green's functions of the central scattering region. The quantities $\Gamma_{L/R}$ are the linewidth functions describing the coupling between the left/right lead and the central region, and can be obtained from $\Gamma_{p}= i(\Sigma^r_p-\Sigma^a_p)$. Here, $\Sigma^{r/a}_p$ is the retarded/advanced
self-energy due to the $p$th semi-infinite lead ($p=L,R$), and can be numerically evaluated using a recursive method.\cite{Rubio}

Before turning on the disorder potential ($W=0$), we know that on each edge there are three conducting channels and the propagation directions of the them are tied to their valley indices. For example, on the upper edge, there are one channel in $K$ valley propagating to the right and two channels in $K'$ valley propagating to the left, as shown schematically in Fig.~\ref{FIG:EdgeState}(b). For the lower edge, the directions of the channels are reversed. Obviously, the two-terminal conductance for the structure should be $G=3$ (in units of $e^2/h$) due to three ballistic transport channels in each direction.

When we increase the disorder strength $W$, backscattering occurs in these edge channels because short-range scatterer can couple the counter-propagating channels at $K$ and $K'$ valleys. This would decrease the conductance. However, at moderate scattering strength, there must be one remaining transport channel as dictated by the total Chern number $\mathcal{C}=-1$. The chirality requires that at the upper edge, this channel propagating to the left, while at the lower edge, it propagates to the right. This should lead to a plateau of $G=1$ for the two-terminal conductance.

\begin{figure}
  \includegraphics[width=8.8cm]{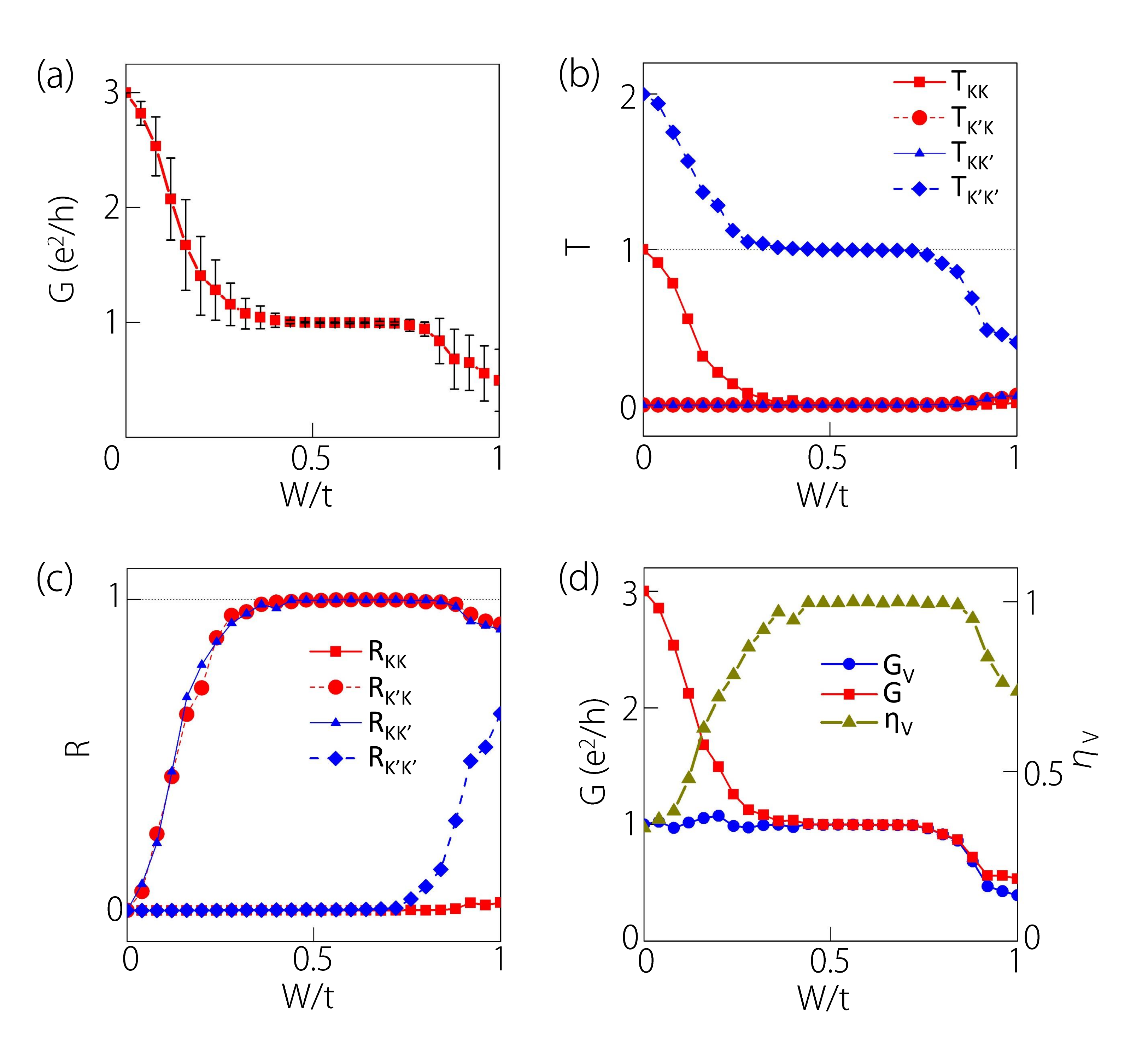}
  \caption{\label{FIG:Conductance} (color online) (a) The two-terminal conductance as a function of disorder strength $W$. Each data point is averaged over 100 disorder configurations. The error bar shows one standard deviation. (b) The valley resolved transmission probability as functions of $W$. (c) The valley resolved reflection probability versus $W$. (d) The charge ($G$), valley ($G_v$) conductance, and the valley polarization ($\eta_V$) as functions of $W$. The model parameters used here are $M=0.5$, $t_{\text{R}_1}=0.045$ and $t_{\text{R}_2}=0.08$. Fermi energy is taken as $E_F=0.004$.}
\end{figure}

In Fig.~\ref{FIG:Conductance}(a) we plot numerical results of the conductance $G$ as a function of the disorder strength $W$. The central scattering region has a width of 480 atomic sites and a length of 1200 atomic sites.
The Fermi level is set at $E_F=0.004$ in the gap and an ensemble of 100 random disorder configurations are taken for each data point. Indeed, as we expected, $G$ starts from the value of $3$ at $W=0$ and decreases with increasing $W$. A quantized plateau of $G=1$ appears around $W=0.5$ with negligibly small fluctuations. This implies that a pair of counter-propagating channels are localized by the short-range scattering and there is one channel left. Further increasing $W$ above $\sim0.75$ eventually destroys the QAH channels by coupling the two chiral channels at the opposite edges through strong scattering across the bulk. All these are consistent with our previous argument.

However, we do not yet know whether the one chiral channel on the $G=1$ plateau still retains a well-defined valley character, although one may intuitively think that after one channel in $K$ valley gets annihilated with one in $K'$ valley, the remaining one should be from $K'$ valley. With finite disorder strength, it is difficult if not impossible to check the valley feature in energy spectrum. Instead, we infer the valley character of the channel from its transport properties. More specifically, we consider the valley resolved transmission probability $T_{vv'}$ ($v, v'\in\{K,K'\}$)
which is defined as the transmission probability from any incoming mode in valley $v'$ of the left lead to any outgoing mode in valley $v$ of the right lead. Hence $T_{vv'}=\sum_{m\in v, n\in v'}T_{mn}$, where $T_{mn}$ is the transmission probability from mode $n$ to mode $m$, $n$ and $m$ label the propagating modes in the left and the right lead respectively. In our case, since we consider the Fermi level in the band gap, the only propagating modes in the leads are the edge modes. The $T_{mn}$ for each pair of incoming and outgoing propagating modes can be calculated using the technique developed in Ref.~\onlinecite{Ando,Kelly,ZhangYY}.

The result for each $T_{vv'}$ as a function of disorder strength is shown in Fig.~\ref{FIG:Conductance}(b). One observes that when $W$ is small ($<0.3$),  both $T_{KK'}$ and $T_{K'K}$ are zero. This is easily understood by inspecting the configuration of channels in Fig.~\ref{FIG:EdgeState}(a). In order for an incoming mode in $K$ valley from the left lead to transfer to an outgoing mode in $K'$ valley at the right lead, it has to cross the insulating bulk hence such probability is negligibly small. For small $W$ ($<0.1$), we approximately have $T_{K'K'}\approx 2T_{KK}$, because the number of channels in $K'$ valley doubles that of the $K$ valley. They both decrease with increasing $W$. When $W$ reaches the plateau region as in Fig.~\ref{FIG:Conductance}(a), $T_{KK}$ vanishes implying that the transport channel at $K$ valley in the central region is totally destroyed. Meanwhile $T_{K'K'}$ shows a quantized plateau at 1. This indicates that the remaining chiral edge channels protected by Chern number $\mathcal{C}=-1$ is of $K'$ valley character. The result suggests the physical picture in which the short-range scattering couples the edge states of the two valleys and destroys a pair of counter-propagating modes (one from each valley), leaving only one edge channel of $K'$ valley in the system. Finally at very large $W$, scattering can couple the two edges, and the plateau is destroyed. The discussed features are schematically shown in Fig.~\ref{FIG:EdgeState}(b) and Fig.~\ref{FIG:EdgeState}(c). At moderate disorder strength (in the plateau region), the central region can be viewed effectively as having one chiral channel in $K'$ valley on each edge [Fig.~\ref{FIG:EdgeState}(c)].

The valley resolved reflection probability $R_{vv'}$ can be defined in a similar way as $T_{vv'}$.\cite{Ando} The results are shown in Fig.~\ref{FIG:Conductance}(c). The key thing to notice is that in the QAH plateau region, $R_{K'K'}$ remains 0 because such reflection process requires the electron to transfer across the insulating bulk to the other edge. This condition, combined with $T_{K'K'}=1$ in this region, means that carriers in $K'$ valley can transmit through the system without reflection while maintaining its valley character. In addition, the rapid increase of $R_{KK'}$ and $R_{K'K}$ at small $W$ demonstrates that the short-range scatterers indeed cause strong backscattering between the counter-propagating channels at each edge.

Based on the valley resolved transmission probability, we could define a valley resolved conductance by
\begin{equation}
G_v=\frac{e^2}{h}\sum_{m;n\in v}T_{mn}, \qquad v\in \{K,K'\},
\end{equation}
which measures the likelihood of transmission of incoming carriers in each valley. Then the total conductance can be written as $G=G_K+G_{K'}$. Analogous to quantities defined for spin transport, we can define a valley conductance $G_V=G_{K'}-G_K$ and the valley polarization $\eta_V=G_V/G$. The numerical results for these quantities are shown in Fig.~\ref{FIG:Conductance}(d). One observes that in the QAH plateau region, the valley polarization shows a plateau of 1, meaning that the transport through the system is fully valley polarized.

From the above results and discussions, we confirm the intuitive picture that we postulated at the beginning of this section. We show that at moderate disorder strength, the scattering localizes a pair of edge channels on each edge, leaving the system in a $\mathcal{C}=-1$ QAH state. The remarkable point is that the remaining one channel still retains its $K'$ valley character. This implies that disorder scattering effectively induces a transition from a VQAH phase with $(\mathcal{C}=-1, \mathcal{C}_v=3)$ to another VQAH phase with $(\mathcal{C}=-1, \mathcal{C}_v=1)$. Such disorder induced VQAH phase with $|\mathcal{C}|=|\mathcal{C}_v|$ always has fully valley-filtered chiral edge channels in the bulk mobility gap, and may be termed as a VQAH Anderson insulator phase, analogous to the concept introduced in the study of transport features of disordered topological insulator systems.\cite{LiJian,Groth} In this case, whether a carrier can be transmitted through the system is determined by its valley index. Hence this phase could be used to realize a perfect valley filter for valleytronics applications.

\section{VQAH phases in bilayer system}

We have shown that VQAH phase can arise in a single layer honeycomb lattice model due to the competition between two types of SOCs. In the following, we show that by combining two such single layers into a bilayer system, more VQAH phases with different $(\mathcal{C},\mathcal{C}_v)$ could be realized. The model we consider is
\begin{equation}
H=H_t+H_b+t_\bot\sum_{\substack{i\in (t, A);\\j\in(b,B)}}(c_i^\dagger c_j+h.c.)+U\sum_i \xi_i c_i^\dagger c_i,
\end{equation}
where $H_t$ and $H_b$ each given by Eq.(\ref{H1}) are the Hamiltonians for the top layer and the bottom layer. The third term on the right hand side is an interlayer coupling. Here we take a bilayer with $AB$-stacking, hence hopping between the nearest $A$ site in the top layer and $B$ site in the bottom layer is considered. The last term is an interlayer bias potential with $\xi_i=\pm 1$ for the two layers.

\begin{figure}
  \includegraphics[width=8cm]{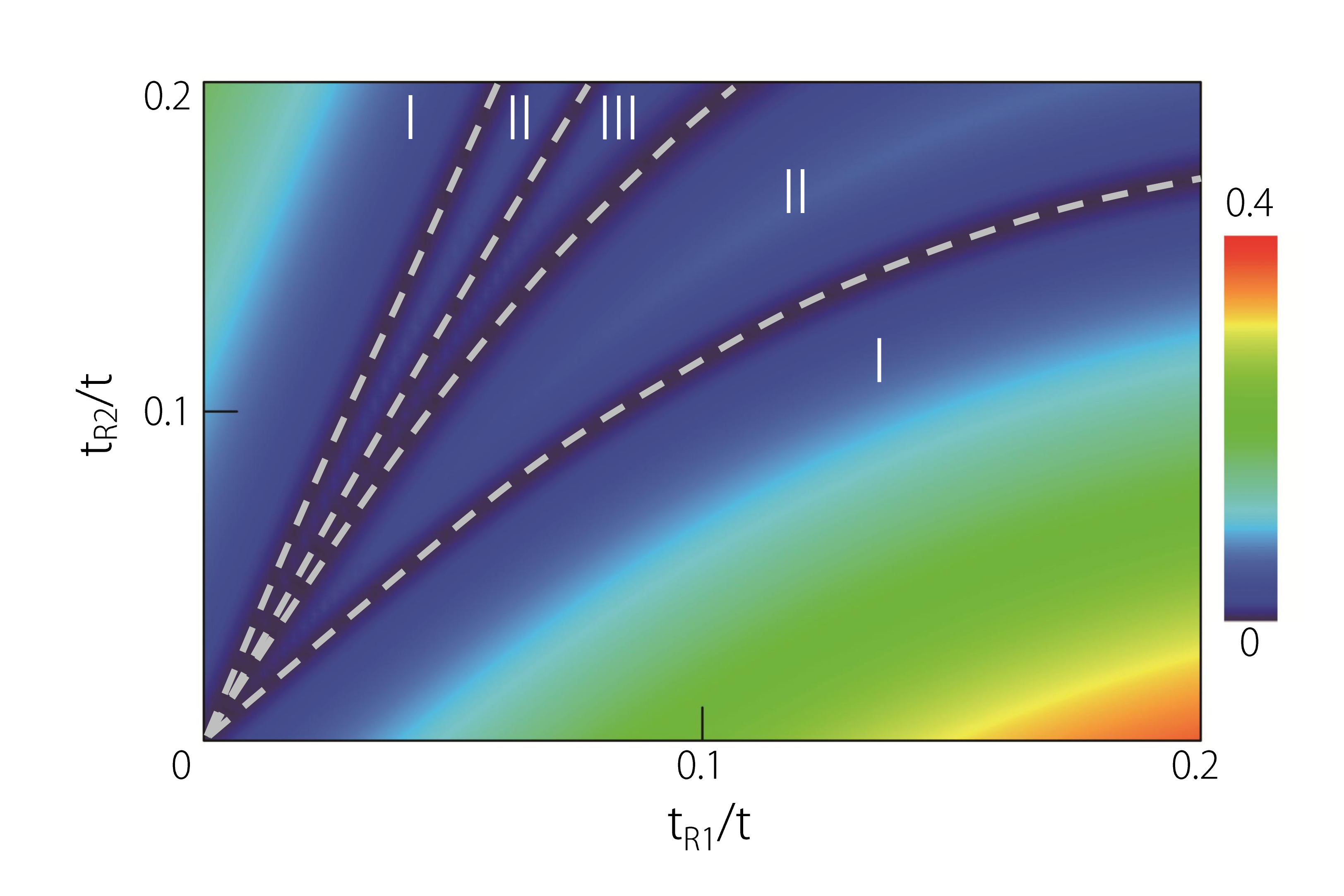}
  \caption{\label{FIG:PhasetR1tR2} (color online) Phase diagram of the bilayer model in the ($t_{\text{R}_1}$, $t_{\text{R}_2}$) plane. Five extended insulating phase regions can be identified. The color indicates the size of the bulk band gap. Other model parameters used here are $M=0.4$, $U=0$, and $t_{\perp}=0.18$.}
\end{figure}

We first set the interlay bias $U=0$. Figure~\ref{FIG:PhasetR1tR2} shows the phase diagram in the ($t_{\text{R}_1}$, $t_{\text{R}_2}$) plane. One observes that there are five phase regions. The color map indicates the magnitude of the bulk band gap. The phase boundaries are the points at which the gap closes. The topological invariants for each phase are listed in Table~\ref{TAB:CherntR1tR2}. It can be seen that Phase I is the conventional QAH phase with $(\mathcal{C}=4, \mathcal{C}_v=0)$ and Phase III is a VQAH phase with $(\mathcal{C}=-2, \mathcal{C}_v=6)$. These two phases can be understood as resulting from the corresponding topological phases in single layer by a direct doubling. Besides these two, interestingly there is one additional phase, Phase II. We find that this phase is also a VQAH phase. It is characterized by $(\mathcal{C}=1, \mathcal{C}_v=3)$, which differs from that of Phase III. Therefore, one sees that by combining two single layer models, it is possible to generate new VQAH phases. The phase transition between each neighboring phase is accompanied by the gap closing and reopening process, which is a general feature of topological phase transitions. In this case, the gap closing only occurs in the $K'$ valley, similar to the single layer case. This is consistent with the variation of the valley topological charge $\mathcal{C}_K$ and $\mathcal{C}_{K'}$ in Table~\ref{TAB:CherntR1tR2}, i.e. $\mathcal{C}_K$ is the same for all the three phases and only $\mathcal{C}_{K'}$ changes.

\newlength\savedwidth
\newcommand\whline{\noalign{\global\savedwidth\arrayrulewidth
                            \global\arrayrulewidth 0.7pt}%
                   \hline \hline
                   \noalign{\global\arrayrulewidth\savedwidth}}

\begin{table}[tbp]
 \caption{\label{TAB:CherntR1tR2} The topological numbers of each phase in Fig.~\ref{FIG:PhasetR1tR2}}
 \begin{tabularx}{8cm}{XXXXX}
  \whline
   Phase  & $\mathcal{C}$ & $\mathcal{C}_v$ & $\mathcal{C}_{K}$ & $\mathcal{C}_{K'}$\\
  \hline
  I   & 4  & 0 &2  & 2  \\
  II  & 1  & 3 &2  & -1 \\
  III & -2 & 6 &2  & -4 \\
  \whline
 \end{tabularx}
\end{table}

\begin{figure}
  \includegraphics[width=8.6cm]{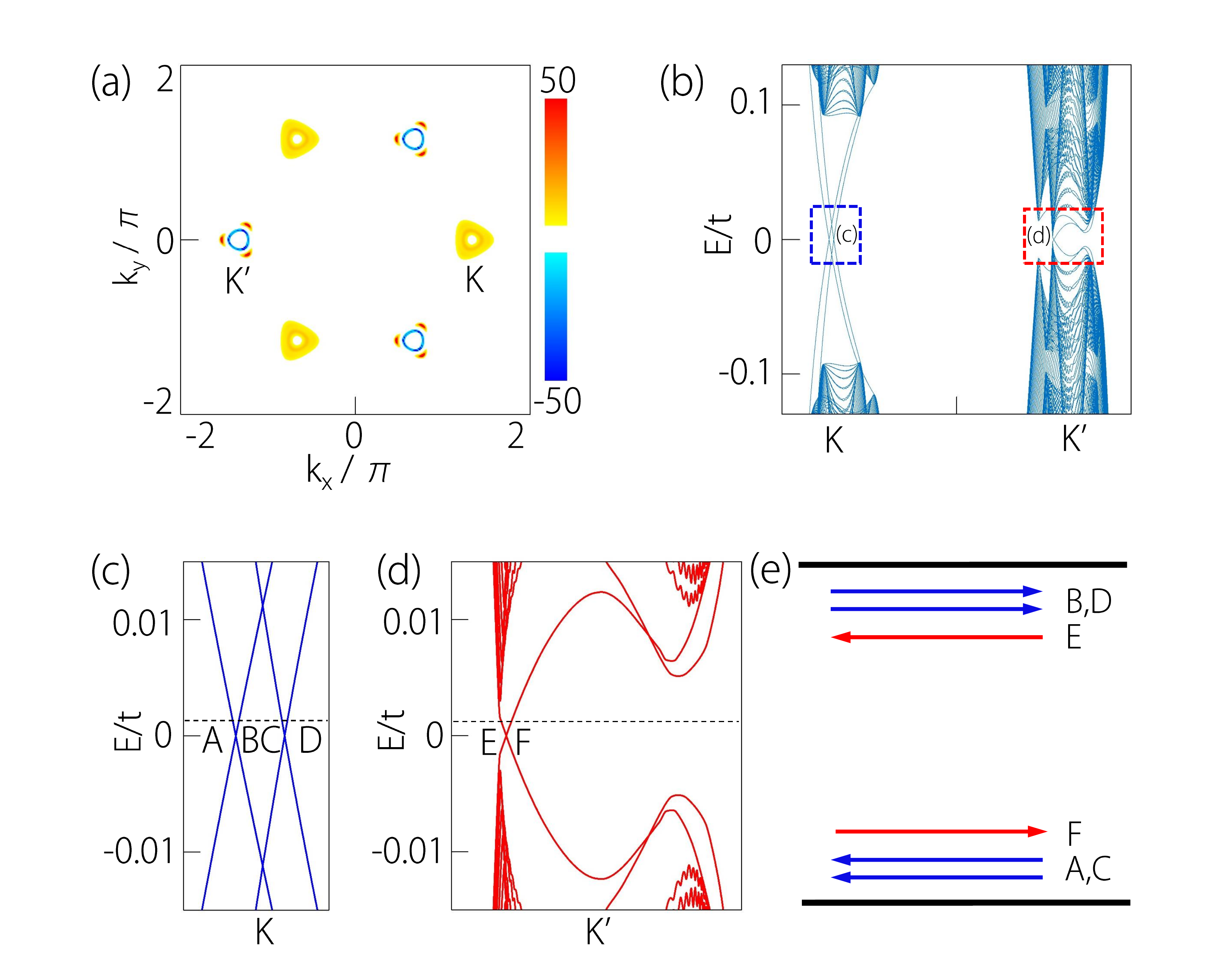}
  \caption{\label{FIG:BerryCp1} (color online) Phase II in Fig.~\ref{FIG:PhasetR1tR2}. (a) Berry curvature distribution in the Brillouin zone. (b) Spectrum of a zigzag-edged ribbon with a width of 400 atomic sites. (c) and (d) The enlarged spectra in the gap region of (b), corresponding to the two valleys $K$ and $K'$. (d) Schematic figure showing the distribution of edge channels labeled in (c) and (d). The parameters are set to be $t_{\text{R}_1}=0.03$, $t_{\text{R}_2}=0.1$, $U=0$, $M=0.4$, and $t_{\perp}=0.18$.}
\end{figure}

\begin{figure}[t]
  \includegraphics[width=8.6cm]{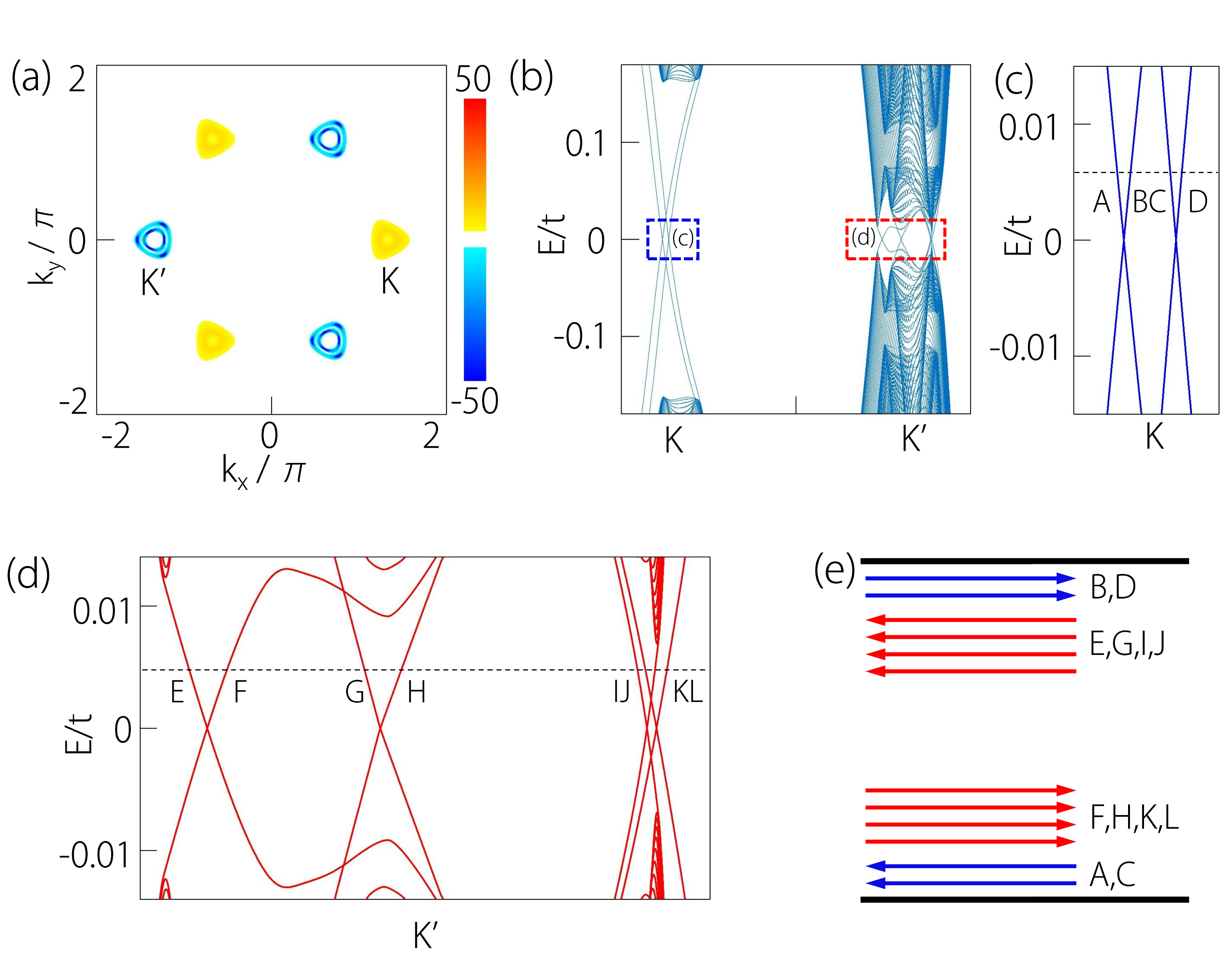}
  \caption{\label{FIG:BerryCn2} (color online) Phase III in Fig.~\ref{FIG:PhasetR1tR2}. (a) Berry curvature distribution in the Brillouin zone. (b) Spectrum of a zigzag-edged ribbon with a width of 400 atomic sites. (c) and (d) The enlarged spectra in the gap region of (b), corresponding to the two valleys $K$ and $K'$. (d) Schematic figure showing the distribution of edge channels labeled in (c) and (d). The parameters are set to be $t_{\text{R}_1}=0.065$, $t_{\text{R}_2}=0.15$, $U=0$, $M=0.4$, and $t_{\perp}=0.18$.}
\end{figure}

In the following, let's have a closer look at the two VQAH phases, Phase II and Phase III. In Fig.~\ref{FIG:BerryCp1}(a) and Fig.~\ref{FIG:BerryCn2}(a) we plot the total Berry curvature distribution $\mathit\Omega(\bm k)$ of the valence bands, which sums over the Berry curvature for each individual valence band. One can find that the nonzero Berry curvatures are mainly concentrated around the valley centers and have an overall opposite sign between the two valleys. This is in contrast to the conventional QAH effect, in which the Berry curvature usually has the same sign for different valleys.\cite{QiaoZH1} For both phases, the asymmetry between the two valleys can be clearly observed. Comparing the two phases, one observes that the curvature distributions at $K$ valley are almost the same, but the curvature at $K'$ valley differ a lot. $\mathit\Omega(\bm k)$ around $K'$ for Phase III has a larger negative magnitude compared with Phase II. This difference leads to the different $\mathcal{C}_{K'}$ between the two phases.

In Fig.~\ref{FIG:BerryCp1}(b), we plot the energy spectra of a zigzag-edged nanoribbon for Phase II. In the zoom-in images in Fig.~\ref{FIG:BerryCp1}(c) and Fig.~\ref{FIG:BerryCp1}(d), one observes that for valley $K$, there are two pairs of gapless edge states while for valley $K'$, there is one pair. On each edge, there are two channels from $K$ valley propagating in one direction and another channel from $K'$ valley propagating in the opposite direction, as shown schematically in Fig.~\ref{FIG:BerryCp1}(e), which is consistent with the bulk topological invariants $(\mathcal{C}=1, \mathcal{C}_v=3)$. This net valley polarization of the edge channels is one signature of VQAH phase.

Similarly, for Phase III, we plot its energy spectrum in Fig.~\ref{FIG:BerryCn2}(b-d), in which one identifies two pairs of gapless edge states in valley $K$ and the other four pairs in valley $K'$. As illustrated in Fig.~\ref{FIG:BerryCn2}(c), on each edge, there are two edge channels from valley $K$ propagating in one direction and four edge channels from valley $K'$ propagating in the opposite direction. Compared with Phase II, the number of channels in $K$ valley remains the same, while the number in $K'$ valley changes from $1$ to $4$. This change reverses the valley polarization of the channels, i.e. now the system has more edge channels in $K'$ than in $K$.

\begin{figure}
  \includegraphics[width=8cm]{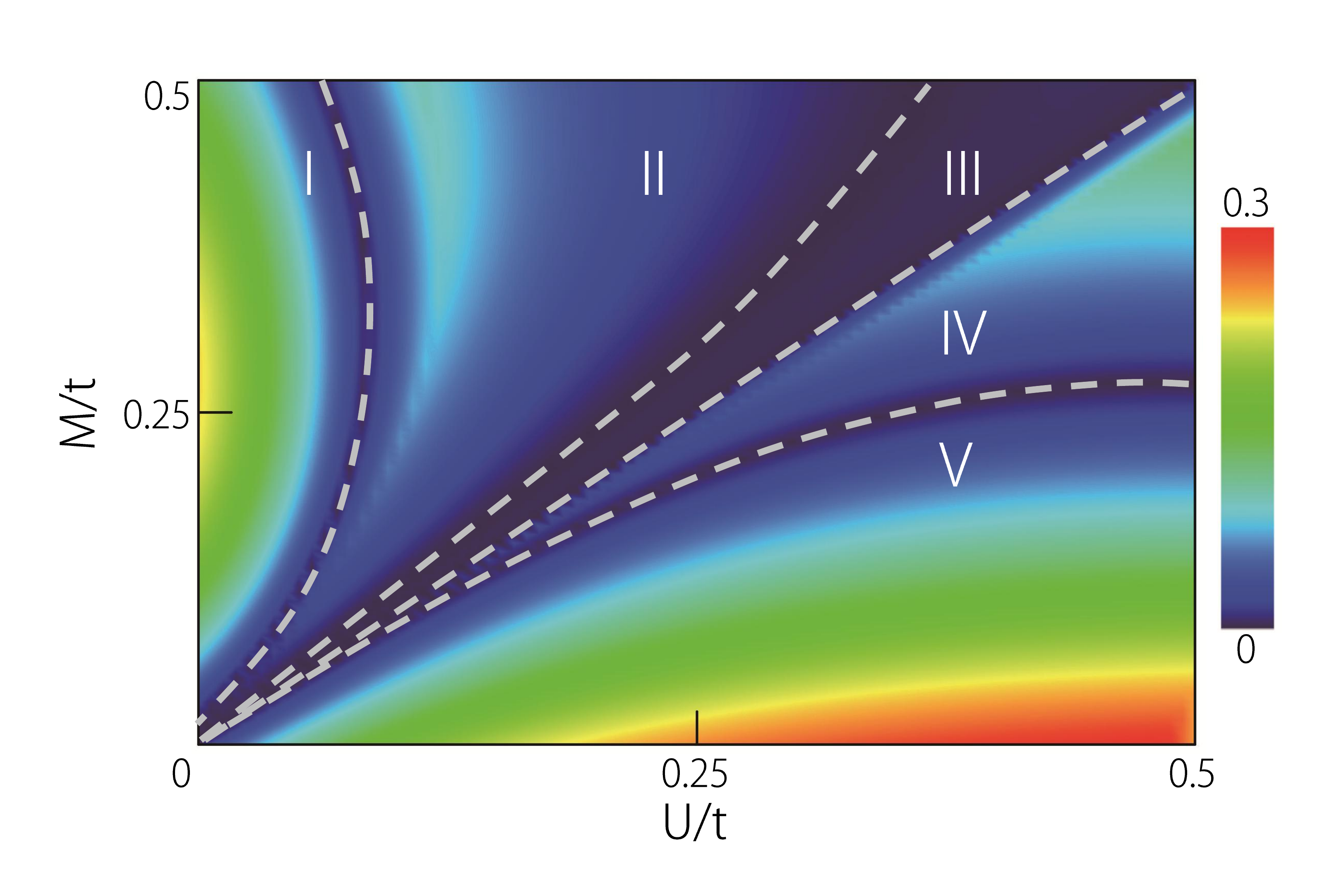}
  \caption{\label{FIG:PhaseUM} (color online) Phase diagram of the bilayer model in the ($U$, $M$) plane. Five extended phase regions can be identified. The color indicate the size of the bulk band gap. Other model parameters are fixed as $t_{\text{R}_1}=0.2$, $t_{\text{R}_2}=0.12$, and $t_{\perp}=0.18$.}
\end{figure}

Since bilayer systems provide another layer degree of freedom, which offers additional controllability. In the following, we examine the effect of tuning interlayer bias potential on the topological phases of the system. In Fig.~\ref{FIG:PhaseUM}, we plot the phase diagram in ($U$, $M$) plane, with fixed SOC strengths. One can identify five different topological phases in this diagram. The topological invariants for each phase are listed in Table~\ref{TAB:ChernMU}. We take the values of $t_{\text{R}_1}$ and $t_{\text{R}_2}$ such that at small $U$, the system is in the conventional QAH phase with $(\mathcal{C}=4, \mathcal{C}_v=0)$ (Phase I in Fig.~\ref{FIG:PhaseUM}). With increasing $U$, we see that the system can undergo topological phase transitions to a series of VQAH phases with $(\mathcal{C}=1, \mathcal{C}_v=3)$ (II), $(\mathcal{C}=-2, \mathcal{C}_v=6)$ (III), $(\mathcal{C}=-3, \mathcal{C}_v=5)$ (IV), and $(\mathcal{C}=0, \mathcal{C}_v=2)$ (V). By inspecting $\mathcal{C}_K$ and $\mathcal{C}_{K'}$, we can see the sequence of gap closing and reopening processes at the two valleys during these phase transitions. For the transition from I to II and from II to III, the gap closing is at $K'$ valley, while for the transition from III to IV, the gap closing happens at $K$ valley. These results are also confirmed by the band structure calculations. Therefore, from above discussion, the transition from conventional QAH to VQAH can also be controlled by the interlayer bias. This is understandable because finite $U$ breaks the inversion symmetry connecting the two valleys, hence could drive the system towards a valley polarized state. Because interlayer potential is generally easier to control in practice, e.g. through gating technique, hence this finding offers a potentially convenient route for engineering a VQAH phase in layered structures. Finally at very large $U$ and small $M$, the system's total Chern number would vanish. The topological charges of the two valleys cancel each other. This is known as quantum valley Hall phase in previous studies.\cite{QiaoZH2}

\begin{table}[htbp]
 \caption{\label{TAB:ChernMU}Chern number contribution of each valley in Fig.~\ref{FIG:PhaseUM}}
 \begin{tabularx}{8cm}{XXXXX}
  \whline
  Phase  & $\mathcal{C}$ & $\mathcal{C}_v$ & $\mathcal{C}_{K}$ & $\mathcal{C}_{K'}$\\
  \hline
  I   & 4  & 0 &2  & 2  \\
  II  & 1  & 3 &2  & -1 \\
  III & -2 & 6 &2  & -4 \\
  IV  & -3 & 5 &1  & -4 \\
  V   & 0  & 2 &1  & -1 \\
  \whline
 \end{tabularx}
\end{table}

\section{Discussion and Summary}

In the analysis of edge channels, the valley index of them is only well-defined provided that the two valleys are separated in momentum when projected to the edge. This depends on the edge orientation. For example, for zigzag edges of a honeycomb lattice, $K$ and $K'$ valleys project to separated points on the edge, which ensures the valley index to be defined. In contrast, for arm-chair edges, the two valleys will be projected to the same point on the edge. Therefore, the edge channels do not have a well-defined valley index and usually strong mixing between them could gap the edge states.\cite{ryce2007} This dependence of the edge states on the edge orientation is analogous to what happens in 3D Dirac and Weyl topological phases.\cite{wan,Rappe1,Fang1,Fang2,YangSY2} Nevertheless, the topological invariants such as $\mathcal{C}$ and $\mathcal{C}_v$ are defined for the bulk, hence do not depend on the edge orientation.

In Sec. III, we used the single layer $(\mathcal{C}=-1,\mathcal{C}_v=3)$ phase as an example to explicitly demonstrate the effects of short-range disorders. Similar physics also happens for other VQAH phases such as the phases that we discussed in the bilayer systems. For example, for the phase
$(\mathcal{C}=-2,\mathcal{C}_v=6)$ as shown in Fig.~\ref{FIG:BerryCn2}, at moderate scattering strength, two pairs of counter-propagating channels would be localized, leaving only two transport channels with $K'$ valley character. Hence the resulting system is equivalent to a $\mathcal{C}=-2$ QAH phase with edge channels fully polarized in valley $K'$. This implies that it is possible to achieve full valley polarized transport with higher conductance values by starting from VQAH states with higher Chern numbers. From the discussion in Sec. III, one also expects that the valley polarized transport plateau could be extended to even higher disorder strength if one could arrange the disorders to distribute more near the edges than in the bulk, thereby suppressing the coupling between the two edges. Increasing the edge roughness may be a possible way to achieve this.

The intrinsic SOC as denoted by $t_\text{SO}$ term is a term that preserves the inversion symmetry, hence is generally not helpful for the purpose of inducing valley polarization. Previous studies have shown that this term could drive the system to a quantum spin Hall phase in the absence of time reversal symmetry breaking.\cite{kane2005a} For the models we considered here, with other parameters fixed, the intrinsic SOC tends to drive the system out of the VQAH phase. However, since the topological phase is protected by the band gap, for small $t_\text{SO}$ such that the gap is not closed, VQAH phase is still maintained. In the single layer model, the inversion symmetry could be broken by a staggered sublattice potential term.\cite{kane2005b} Similar to the effect of interlayer bias potential for the bilayer model, this term could generate valley polarization, and is capable of driving the system from conventional QAH phase to the VQAH phase in single layer model.

Finally, a particular lattice model is adopted here for the proposition and the study of the novel VQAH phase. We emphasize that the essential features we discuss here, such as the valley-polarization of the gapless edge channels and the disorder effects on the edge channels, are general features of the VQAH phase and are not particular to one specific model. In reality, several 2D materials with low buckled honeycomb lattice structure have been discovered or proposed.\cite{xu2013,zhu2014,song2014,cliu2014} In principle, the strength of each individual terms in our model Hamiltonian Eq.(\ref{H1}) could be induced and controlled. For example, the extrinsic Rashba SOC could be generated by substrate or adsorbed atoms.\cite{Dedkov,Varykhalovet,bala} The exchanged coupling could be generated by defects, magnetic dopants, or in proximity to a magnetic insulator.\cite{DingJ2,TEelbo,WeiP,wang2009} Intrinsic SOC and intrinsic Rashba SOC could be controlled by structural deformation through applied strains.\cite{LiuCC2} Furthermore, the candidate material is not limited to those with 2D honeycomb lattice structure. Any multi-valley systems are possible. Therefore, although fine-tuning the various parameters to achieve the VQAH phase is a challenging task, with the advance in discovering new 2D materials and in developing new technique to control interactions at submicron scale, as demonstrated in the recent realization of QAH phase, it is promising to also achieve the fascinating VQAH phase in the future.

In summary, we investigated in detail the novel VQAH phase in single layer and bilayer systems. We provide a clear physical picture of the topological phase transition from conventional QAH phase to the VQAH phase. We studied the transport properties of the edge channels. With short-range disorders, pairs of counter-propagating edge channels (one from each valley in a pair) could be destroyed. However, at moderate scattering strength, the transport coefficients exhibit a plateau on which the transport is fully valley-filtered, leading to a VQAH Anderson insulator phase. This remarkable effect could be used for designing valley filters for valleytronic applications. Much richer phase diagrams are shown for the bilayer system with multiple VQAH phases. We demonstrate the controllability of the topological phase transition by tuning the system parameters, especially the interlayer bias potential. The study presented here endows the valley transport with topological protection, which is very important for realizing robust performance of information processing based on valley degree of freedom.
\\

\emph{Acknowledgement.} The authors would like to thank Jian Li, Zhenhua Qiao, Y.-W. Wang, and D. L. Deng for helpful discussions. This work was supported by the SUTD-SRG-EPD2013062, the NSFC (Grants Nos. 11174022, 11174337, 11225418, and 11374219), the MOST Project of China (Grants Nos. 2014CB920901, 2014CB920903, 2011CBA00100), and the SRFDPHE of China (No. 20121101110046).


\begin{thebibliography}{999}

\bibitem{Laughlin} R. B. Laughlin, Phys. Rev. B \textbf{23}, 5632 (1981).

\bibitem{Thouless} D. J. Thouless, M. Kohmoto, M. P. Nightingale, and M. den Nijs, Phys. Rev. Lett. \textbf{49}, 405 (1982).

\bibitem{hald} F. D. M. Haldane, Phys. Rev. Lett. {\bf 61}, 2015 (1988).
\bibitem{onod} M. Onoda and N. Nagaosa, Phys. Rev. Lett. {\bf 90}, 206601 (2003).
\bibitem{qi} X. L. Qi, Y. S. Wu, and S. C. Zhang, Phys. Rev. B {\bf 74}, 085308 (2006).
\bibitem{qi2008} X. L. Qi, T. L. Hughes, and S. C. Zhang, Phys. Rev. B {\bf 78}, 195424 (2008).
\bibitem{wang2014} J. Wang, B. Lian, and S. C. Zhang, Phys. Rev. B {\bf 89}, 085106 (2014).

\bibitem{YuR} R. Yu, W. Zhang, H.-J. Zhang, S.-C. Zhang, Xi Dai, Zhong Fang, Science \textbf{329}, 61 (2010).
\bibitem{ChangCZ} C. Z. Chang \textit{et al.}, Science \textbf{340}, 167 (2013).
\bibitem{chec} J. G. Checkelsky, R. Yoshimi, A. Tsukazaki, K. S. Takahashi, Y. Kozuka, J. Falson, M. Kawasaki, and Y. Tokura, Nat. Phys. \textbf{10}, 731 (2014).
\bibitem{kou} X. Kou \emph{et al.}, arXiv:1406.0106.
\bibitem{JiangH} H. Jiang, Z. Qiao, H. Liu, and Q. Niu, Phys. Rev. B {\bf 85}, 045445 (2012).

\bibitem{KaneCL1} M. Z. Hasan and C. L. Kane, Rev. Mod. Phys. \textbf{82}, 3045 (2010).
\bibitem{ZhangSC1} X.-L. Qi and S.-C. Zhang, Rev. Mod. Phys. \textbf{83}, 1057 (2011).

\bibitem{ryce2007} A. Rycerz, J. Tworzydlo, and C. W. J. Beenakker, Nat. Phys. \textbf{3}, 172 (2007).
\bibitem{guna2006} O. Gunawan, Y. P. Shkolnikov, K. Vakili, T. Gokmen, E. P. De Poortere, and M. Shayegan, Phys. Rev. Lett. {\bf 97}, 186404 (2006).
\bibitem{xiao2007} D. Xiao, W. Yao, and Q. Niu, Phys. Rev. Lett. {\bf 99}, 236809 (2007).
\bibitem{yao2008}  W. Yao, D. Xiao, and Q. Niu, Phys. Rev. B {\bf 77}, 235406 (2008).
\bibitem{zhu2012}  Z. Zhu, A. Collaudin, B. Fauque, W. Kang, and K. Behnia, Nat. Phys. {\bf 8}, 89 (2012).
\bibitem{xu2014}   X. Xu, W. Yao, D. Xiao, and T. F. Heinz, Nat. Phys. {\bf 10}, 343 (2014).
\bibitem{KFMak} K. F. Mak, K. L. McGill, J. Park, P. L. McEuen, Science, \textbf{334}, 1489 (2014).

\bibitem{Morpurgo} I. Martin, Ya. M. Blanter, and A. F. Morpurgo, Phys. Rev. Lett. \textbf{100}, 036804 (2008).
\bibitem{YaoW2} W. Yao, S. A. Yang, and Q. Niu, Phys. Rev. Lett. \textbf{102}, 096801 (2009).
\bibitem{QiaoZH1} Z. Qiao, S. A. Yang, W. Feng, W.K. Tse, J. Ding, Y. Yao, J. Wang, and Q. Niu , Phys. Rev. B \textbf{82}, 161414(R) (2010).
\bibitem{QiaoZH2} Z. Qiao, W.K. Tse, H. Jiang, Y. Yao, and Q. Niu, Phys. Rev. Lett. \textbf{107}, 256801 (2011).
\bibitem{MacDonald2} F. Zhang, J. Jung, G. A. Fiete, Q. Niu, and A. H. MacDonald, Phys. Rev. Lett. \textbf{106}, 156801 (2011).
\bibitem{YKim} Y. Kim, K. Choi, and J. Ihm, Phys. Rev. B \textbf{89}, 085429 (2014).

\bibitem{PanH} H. Pan, Z. Li, C.-C. Liu, G. Zhu, Z. Qiao, and Y. Yao, Phys. Rev. Lett. \textbf{112}, 106802 (2014).

\bibitem{Lalmi} B. Lalmi, H. Oughaddou, H. Enriquez, A. Kara, S. Vizzini, B. Ealet, and B. Aufray, Appl. Phys. Lett. \textbf{97}, 223109 (2010).
\bibitem{LiuCC1} C. C. Liu, W. Feng, and Y. Yao, Phys. Rev. Lett. \textbf{107}, 073802 (2011).
\bibitem{LiuCC2} C. C. Liu, H. Jiang, and Y. Yao, Phys. Rev. B \textbf{84}, 195430 (2011).
\bibitem{Vogt} P. Vogt, P. D. Padova, C. Quaresima1, J. Avila, E. Frantzeskakis, M.C. Asensio, A. Resta, B. Ealet, and G. L. Lay,  Phys. Rev. Lett. \textbf{108}, 155501 (2012).
\bibitem{Fleurence} A. Fleurence, R. Friedlein, T. Ozaki, H. Kawai, Y. Wang, and Y. Yamada-Takamura, Phys. Rev. Lett. \textbf{108}, 245501 (2012).
\bibitem{Ezawa1} M. Ezawa, Phys. Rev. Lett. \textbf{109}, 055502 (2012).
\bibitem{ChenL} L. Chen, C. C. Liu, B. Feng, X. He, P. Cheng, Z. Ding, S. Meng, Y. Yao, and K. Wu , Phys. Rev. Lett. \textbf{109}, 056804 (2012).
\bibitem{Tsai} W. F. Tsai, C. Y. Huang, T. R. Chang, H. Lin, H. T. Jeng, and A. Bansil, Nat. Commun. \textbf{4}, 1500 (2013).

\bibitem{kane2005a} C. L. Kane and E. J. Mele, Phys. Rev. Lett. {\bf 95}, 226801 (2005).
\bibitem{kane2005b} C. L. Kane and E. J. Mele, Phys. Rev. Lett. {\bf 95}, 146802 (2005).

\bibitem{YangSY1} S. A. Yang, H. Pan, and F. Zhang, arXiv:1409.1977.
\bibitem{PanH2} H. Pan, X. Li, Z. Qiao, C.-C. Liu, Y. Yao, and S. A. Yang, New J. Phys. {\bf 16}, 123015 (2014).

\bibitem{naga2010} N. Nagaosa, J. Sinova, S. Onoda, A. H. MacDonald, and N. P. Ong, Rev. Mod. Phys. {\bf 82}, 1539 (2010).

\bibitem{volovik}
 G. E. Volovik, {\it The Universe in a Helium Droplet} (Clarendon Press, Oxford, 2003); JETP Lett. {\bf 93}, 66 (2011).

\bibitem{Datta} S. Datta, \emph{Electronic Transport in Mesoscopic Systems} (Cambridge University Press, Cambridge, UK, 2003).

\bibitem{Rubio} M. P. L\'{o}pez-Sancho, J. M. L\'{o}pez-Sancho, and J. Rubio, J. Phys. F \textbf{14}, 1205 (1984); \textbf{15}, 851 (1985).

\bibitem{Ando} T. Ando, Phys. Rev. B \textbf{44}, 8017 (1991);
\bibitem{Kelly} P. A. Khomyakov, G. Brocks, V. Karpan, M. Zwierzycki, and P. J. Kelly, Phys. Rev. B \textbf{72}, 035450 (2005).

\bibitem{ZhangYY} Y.-Y. Zhang, M. Shen, X.-T. An, Q.-F. Sun, X.-C. Xie, K. Chang, and S.-S. Li, Phys. Rev. B \textbf{90}, 054205 (2014).

\bibitem{LiJian} J. Li, R.-L. Chu, J. K. Jain, and S.-Q. Shen, Phys. Rev. Lett. {\bf 102}, 136806 (2009).
\bibitem{Groth} C. W. Groth, M. Wimmer, A. R. Akhmerov, J. Tworzydlo, and C. W. J. Beenakker, Phys. Rev. Lett. {\bf 103}, 196805 (2009).


\bibitem{wan} X. Wan, A. Turner, A. Vishwanath, and S. Y. Savrasov, Phys. Rev. B {\bf 83}, 205101 (2011).
\bibitem{Rappe1} S. M. Young, S. Zaheer, J. C. Y. Teo, C. L. Kane, E. J. Mele, and A. M. Rappe, Phys. Rev. Lett. {\bf 108}, 140405 (2012).
\bibitem{Fang1} Z. Wang, Y. Sun, X. Chen, C. Franchini, G. Xu, H. Weng, X. Dai, and Z. Fang, Phys. Rev. B {\bf 85}, 195320 (2012).
\bibitem{Fang2} Z. Wang, H. Weng, Q. Wu, X. Dai, and Z. Fang, Phys. Rev. B {\bf 88}, 125427 (2013).
\bibitem{YangSY2} S. A. Yang, H. Pan, and F. Zhang, Phys. Rev. Lett. \textbf{113}, 046401 (2014).

\bibitem{xu2013} Y. Xu, B. Yan, H.-J. Zhang, J. Wang, G. Xu, P. Tang, W. Duan, and S.-C. Zhang, Phys. Rev. Lett. {\bf 111}, 136804 (2013).
\bibitem{zhu2014} Z. Zhu and D. Tomanek, Phys. Rev. Lett. {\bf 112}, 176802 (2014).
\bibitem{song2014} Z. Song, C.-C. Liu, J. Yang, J. Han, M. Ye, B. Fu, Y. Yang, Q. Niu, J. Lu, and Y. Yao, arXiv:1402.2399.
\bibitem{cliu2014} C.-C. Liu, S. Guan, Z. Song, S. A. Yang, J. Yang, and Y. Yao, Phys. Rev. B {\bf 90}, 085431 (2014).


\bibitem{Dedkov} Yu. S. Dedkov, M. Fonin, U. R¨¹diger, and C. Laubschat, Phys. Rev. Lett. \textbf{100}, 107602 (2008).
\bibitem{Varykhalovet} A. Varykhalov \emph{et al.}, Phys. Rev. Lett. \textbf{101}, 157601 (2008).
\bibitem{bala} J. Balakrishnan, G. K. W. Koon, M. Jaiswal, A. H. Castro Neto, and B. Ozyilmaz, Nat. Phys. {\bf 9}, 284 (2013).

\bibitem{DingJ2} J. Ding, Z. H. Qiao, W. X. Feng, Y. G. Yao, and Q. Niu, Phys. Rev. B \textbf{84}, 195444 (2011).

\bibitem{TEelbo} T. Eelbo, M. Wasniowska, P. Thakur, M. Gyamfi, B. Sachs, T. O. Wehling, S. Forti, U. Starke, C. Tieg, A. I. Lichtenstein, and R. Wiesendanger, Phys. Rev. Lett. \textbf{110}, 136804 (2013).

\bibitem{WeiP} P. Wei, F. Katmis, B. A. Assaf, H. Steinberg, P. JarilloHerrero, D. Heiman, and J. S. Moodera, Phys. Rev. Lett. \textbf{110}, 186807 (2013).

\bibitem{wang2009} Y. Wang, Y. Huang, Y. Song, X. Zhang, Y. Ma, J. Liang, and Y. Chen, Nano Lett. {\bf 9}, 220 (2009)


%


\end{thebibliography}
\end{document}